\documentclass[11pt]{article}

\usepackage[final]{acl}

\usepackage{times}
\usepackage[listings]{tcolorbox}
\usepackage{latexsym}
\usepackage{amsmath}
\usepackage{amssymb}
\usepackage{booktabs}
\usepackage{multirow}
\usepackage[T1]{fontenc}
\usepackage[utf8]{inputenc}

\usepackage{microtype}

\usepackage{inconsolata}

\usepackage{graphicx}
\usepackage{fontawesome5}

\newcommand{\projecturl}{https://sphere-music-binauralization.github.io}  % TODO: wire in real URL

\title{SPHERE: Automatic Music Upmixing via Audio Language Model Post-Training with Spatial Heuristic Rewards}

\author{
  \textbf{Zixun Guo\textsuperscript{1,2}}\thanks{Work done at Meta.},
  \textbf{Calvin Murdock\textsuperscript{1}},
  \textbf{Sanjeel Parekh\textsuperscript{1}},
  \textbf{W. Owen Brimijoin\textsuperscript{1}},
\\
  \textbf{Simon Dixon\textsuperscript{2}},
  \textbf{Joshua Reiss\textsuperscript{2}},
  \textbf{Ishwarya Ananthabhotla\textsuperscript{1}}
\\
\\
  \textsuperscript{1}Meta Reality Labs, USA
  \textsuperscript{2}Queen Mary University of London, UK
\\
 \small{
   \faGlobe~\href{\projecturl}{Project Demo Website}
 }
}

\begin{document}
\maketitle
\begin{abstract}

In this paper, we study the task of automatic music upmixing, wherein a system predicts spatial mixing parameters from a multi-stem recording. Different from existing methods that rely on task-specific music encoders, we approach this task via audio language model (ALM) post-training, leveraging rich representations from existing ALMs, which encode both music semantics and mixing knowledge. Specifically, we propose a post-training recipe that first employs rejection sampling SFT, followed by reinforcement learning (RL) with verifiable rewards (RLVR) via GRPO.  We propose \textsc{Sphere} (Spatial Heuristic Rewards), a deterministic reward suite inspired by music mixing conventions, to guide our post-training. It consists of 6 perceptually-motivated sub-rewards and encourages the output mix to be centered, balanced and spacious. 
More broadly, our results suggest that expert domain knowledge can be encoded as verifiable rewards and distilled into language models, without task-specific architectures.% Through analyzing the contribution of each pair of sub-rewards during RL, we find that the sub-rewards in \textsc{Sphere} form a non-linear reward landscape, naturally resistant to reward hacking.
\end{abstract}

\section{Introduction}
\label{sec:intro}

Automatic music mixing is a well-studied task wherein a system receives isolated music stems and produces a music mix \cite{deman2017ten}, inferring mixing parameters for each stem (gain, panning, etc.).
In this paper, we address the task of automatic binaural upmixing, which extends the automatic music mixing task to the spatial domain \cite{begault19943dsound}. In particular, we additionally seek to predict spatial positioning parameters of each stem (i.e., azimuth, elevation) to create a musical presentation that appears three-dimensional and immersive in nature. 

Existing deep learning based music mixing systems usually rely on task-specific encoders (e.g., CNN) to extract music representations to infer mixing parameters implicitly or explicitly \cite{steinmetz21dmcmix, ramirez22outofdomain, Lee24MixingGraphs}.
We argue that these encoders are limited in their ability to encode musical semantics, such as the instrument grouping of stems, which plays a key role in a mix engineer's mixing process.
% such as recognizing musical roles of music stems, which music mixing engineers would rely on normally.
\begin{figure}
\centering
\includegraphics[width=\columnwidth]{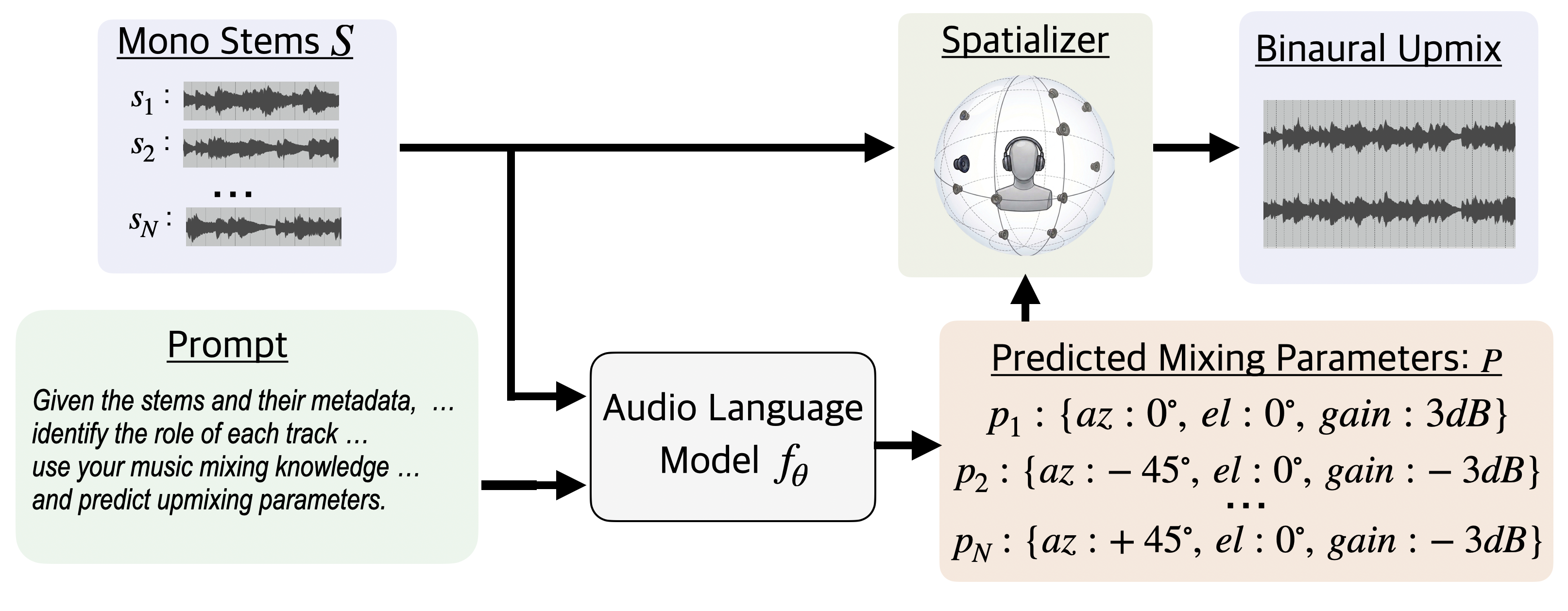}
\caption{Automatic music upmixing via audio language modeling. Az. and El. stand for azimuth and elevation.}
\label{fig:overview}
\end{figure}

Meanwhile, recent Audio Language Models (ALMs) have shown superb audio understanding capabilities \cite{chu2023qwenaudio, xu2025qwen25omni, gemini2.5, ghosh2025musicflamingo}. Since ALMs are usually also trained with internet-scale
text, these models might have music mixing knowledge encoded \cite{doh2025llmfx}. Hence, it is reasonable to believe when prompted with multitrack recording stems with mixing instructions,
these models should be able to predict music mixing parameters, as shown in Figure \ref{fig:overview}. Indeed, our preliminary study shows
that frontier ALMs understand music semantics in multi-track recordings and can produce decent mixing parameters zero-shot. This motivates us to tackle automatic music upmixing via ALMs, leveraging their rich pretrained representations.
However, zero-shot ALMs do not produce mixes with consistent quality. We argue this can be addressed through supervised fine-tuning (SFT) and reinforcement learning (RL) post-training.
% Indeed, our preliminary study shows
% that frontier ALMs understand music semantics in multi-track recordings and can produce decent mixing parameters zero-shot. 
% This motivates us to tackle automatic music upmixing via ALMs, leveraging their rich pretrained representations.
% However, zero-shot ALMs do not always produce mixes with consistent quality. We argue this can be addressed through supervised fine-tuning (SFT) and reinforcement learning (RL) post-training.

In this paper, we approach automatic binaural music upmixing via audio language model post-training, guided by \textsc{Sphere} (Spatial Heuristic Rewards), a deterministic reward suite we propose that draws
inspiration from music mixing conventions, computed as follows. We first compute six perceptually-motivated raw metrics measuring centering, spread, and level balance of the mix.
We then derive data-driven thresholds and apply shaping functions to convert raw
metrics into continuous sub-rewards, which are then summed. \textsc{Sphere} serves two roles: First, we utilize \textsc{Sphere} to filter high-quality outputs from a teacher model and conduct SFT on a smaller student model. We then apply RL via GRPO \cite{shao24deepseekmath} to further optimize the student against
\textsc{Sphere}, enabling it to explore better spatial parameters.

Our subjective listening test (A/B Preference) with 44 participants shows that high-reward mixes are preferred in 76.8\% of trials, confirming \textsc{Sphere} is a reliable proxy for human spatial audio preference. Moreover, after post-training, objective results on the MedleyDB dataset \cite{bittner2014medleydb} show that our model surpasses frontier ALMs by a large margin on \textsc{Sphere}, demonstrating the effectiveness of the post-training pipeline. Last but not least, our ablation study shows that the combination of sub-rewards in \textsc{Sphere} creates a reward landscape that cannot be trivially maximized: removing any pair leads to reward hacking. Only when all six components are present does RL training remain stable, achieving the highest total reward with the lowest variance.

% We aim to answer the following research questions: 1. Does \textsc{Sphere} align with human spatial audio preference? 2. Can post-training with \textsc{Sphere} enable a small student model to match or surpass a much larger teacher?
% 3. If so, what is each sub-reward's contribution during SFT and RL? Our subjective listening test (A/B Preference) with 44 participants shows that high-reward mixes are preferred in 76.8\% of trials,
% confirming \textsc{Sphere} is a reliable proxy for human spatial audio preference. Moreover, after post-training, objective results on the MedleyDB dataset \cite{bittner2014medleydb} show that our model surpasses frontier ALMs
% by a large margin on \textsc{Sphere}, demonstrating the effectiveness of the post-training pipeline.
% Last but not least, our ablation study shows that the combination of sub-rewards in \textsc{Sphere} creates a reward landscape that cannot be trivially maximized: removing any pair leads to varying degrees of reward hacking.
% % Only when all six components are present does RL training remain stable, achieving the highest total reward with the lowest variance.

In summary, our contributions are as follows: 
\begin{enumerate}
    \item To the best of our knowledge, we are the first to tackle automatic music upmixing via ALM post training, leveraging rich music understanding capabilities in pretrained ALMs.
% This offers full explainability and reasoning traces where users can interact with the predicted mixing parameters.
    \item We propose \textsc{Sphere} (Spatial Heuristic Rewards), a reward function inspired by music mixing conventions, that encourages the final mix to be centered, balanced, and to have a wide spatial impression. Through a subjective listening test, we empirically show \textsc{Sphere} aligns with human preference and prevents reward hacking during RL.
    \item We show that our post-trained model outperforms frontier ALMs on our reward suite, which also serves as our evaluation metric.
  \item More broadly, our results suggest that expert domain knowledge can be
  encoded as verifiable rewards and distilled into a language model through
  post-training. This offers a practical recipe for data-limited domains,
  potentially beyond music mixing.
\end{enumerate}

%     \item Our post-trained model, despite being much smaller, outperforms frontier ALMs on \textsc{Sphere}.
% % with RL further improving over SFT on the MedleyDB dataset.

\section{Related Work}
\label{sec:related}
% \ishwarya{The majority of this section as it stands should go to the introduction (describing the challenges of music mixing, the promise of using ALMs for music content analysis and mixing, when coupled with post-training).The Focus Related Work on concrete SOTA approaches to binaural upmixing.  Rewrite.}

Existing deep learning approaches to automatic music mixing typically follow an encoder-decoder design, where an encoder
extracts latent music representations, and a decoder predicts either interpretable mixing parameters \cite{steinmetz21dmcmix, Vanka24DiffMST, Lee24MixingGraphs} or fully-mixed audio directly \cite{ramirez22outofdomain, Moliner25Megami}.
For binaural upmixing, prior work has additionally relied on other modalities to learn mono or stereo to binaural conversion \cite{gao2019visualsound, LluisCH22point2sound}, or predict spatial audio
directly from stereo inputs, inferring the spatial cues from stereo inputs implicitly  
\cite{Grundhuber24NBU, yang22upmixviastyletransfer, zang2024ambisonizer}.
% These systems face several limitations: systems that predict parameters offer more explicit user control, while end-to-end systems lack interpretability and limit further user interaction (e.g., adjusting mixing parameters).
In most cases, these methods have limited semantic understanding of multi-stem recordings or mixing conventions as the encoders are mostly CNN-based and trained from scratch, except \citet{Moliner25Megami}, who use CLAP embeddings.

Meanwhile, recent audio language models (ALMs) have demonstrated strong audio and music understanding capabilities \cite{chu2023qwenaudio, xu2025qwen25omni, gemini2.5}.
Some ALMs can process multiple audio files (e.g., multitrack music recordings) within a single prompt and produce descriptive text, making them suited for music mixing.
However, general-purpose ALMs are not trained on spatial mixing data, resulting in inconsistent zero-shot performance.
To improve task-specific performance, post-training via SFT then RL has proven effective for aligning LLM outputs with desired behaviors, via verifiable rewards \cite{shao24deepseekmath, hernandez2026text} or human preferences \cite{cideron24MusicRL}.
Yet applying post-training to music upmixing requires a reward function that captures perceptual mixing quality and aligns with general listener preference. In this work, we aim to design such a reward function, drawing inspiration from music mixing conventions \cite{pestana14bestpractices,lopes2023instrument}.

\section{Method}
\label{sec:method}

\begin{figure*}[t]
\centering
\includegraphics[width=\textwidth]{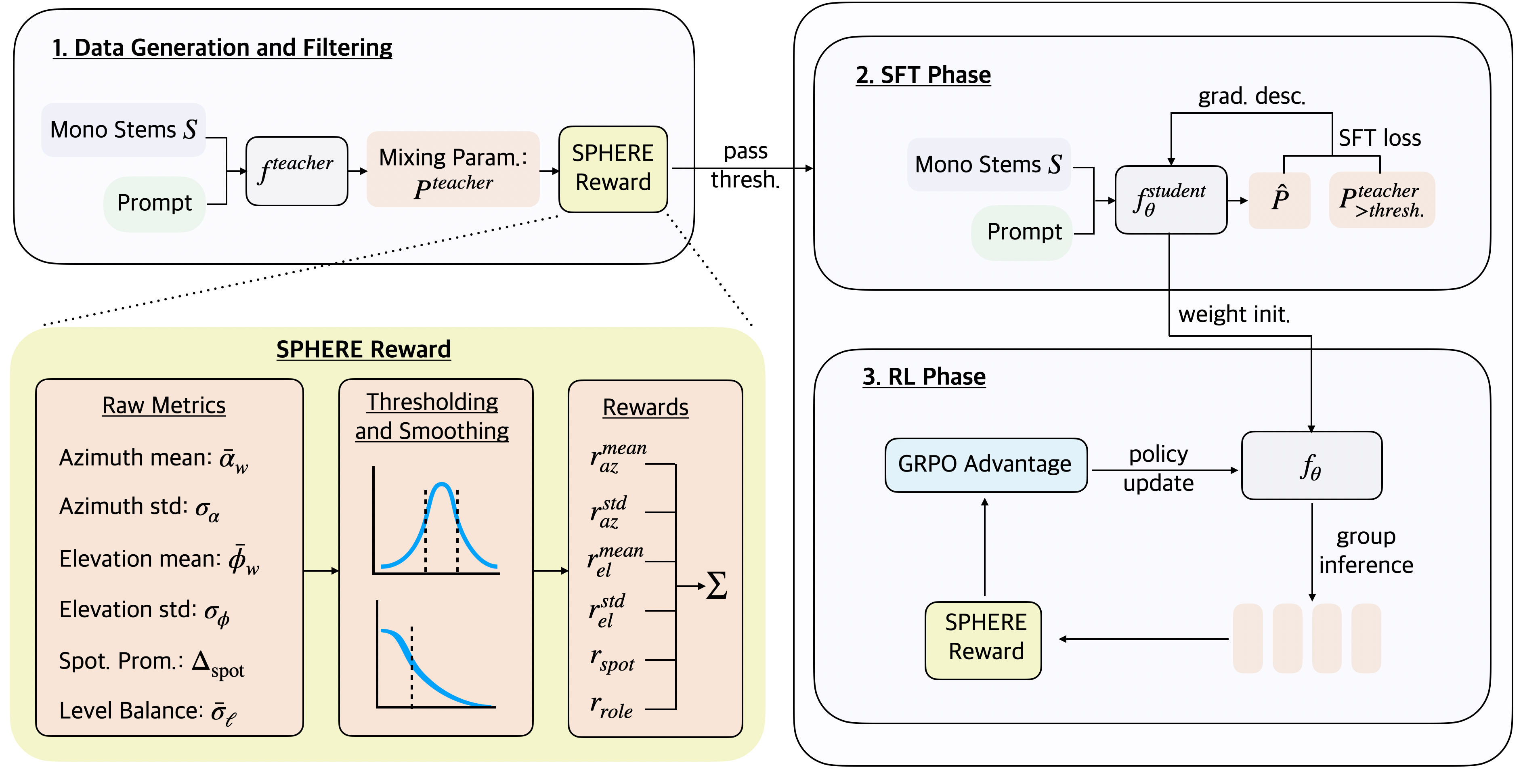}
\caption{Overall post-training paradigm. 1. A teacher model generates spatial mixing parameters, which are filtered by $R_{\textsc{Sphere}}$. 2. SFT finetunes a student on the filtered data. 3. RL further optimizes the student using $R_{\textsc{Sphere}}$.}
\label{fig:paradigm}
\end{figure*}
% $R_{\textsc{Sphere}}$ (bottom-left) computes six raw spatial metrics, applies data-driven thresholding and reward shaping, and sums the resulting continuous sub-rewards.}

\subsection{Task Formulation}
\label{sec:task}

 The overall inference pipeline is shown in Figure~\ref{fig:overview}. Let $\mathcal{S} = \{s_i\}_{i=1}^{N}$ denote the $N$ mono stems of a
  multitrack recording, and $T$ denote a text prompt that includes mixing instructions \cite{deman2019intelligent, pestana14bestpractices} as well as other descriptions of the stems (e.g., metadata, loudness in LUFS). We aim to learn a function
  $f_{\theta}(\mathcal{S}, T)$ that maps $\mathcal{S}$ and $T$ to a set of spatial parameters $P = \{p_i\}_{i=1}^{N}$, where $p_i = (\alpha_i, \phi_i, g_i)$ denotes
  the predicted azimuth, elevation (in degrees), and gain (in dB) of stem $s_i$. In practice, $f_\theta$ is implemented as an ALM that
  receives the mono stems as audio tokens interleaved with tokenized $T$, and produces $P$ as JSON outputs. A binaural spatializer then applies the predicted gain $g_i$ to each stem $s_i$ and places it at position $(\alpha_i, \phi_i)$ by convolving with a corresponding proprietary HRTF, producing a two-channel binaural mix for headphone playback.

% Our objective is to create an upmix that achieves a wide spatial impression while maintaining level balance across stems.

\subsection{Overall Post-Training Paradigm}
\label{sec:paradigm}
The post-training paradigm is shown in Figure \ref{fig:paradigm}.
To acquire paired data for SFT, a teacher model $f_{\mathrm{teacher}}$ generates candidate spatial parameters for each multitrack recording. To ensure data quality, we score each teacher output with $R_{\textsc{Sphere}}$ (to be introduced in Section~\ref{sec:reward_design}) and we retain data points that exceed a quality threshold $\kappa$. A student model $f_\theta$ is then finetuned on the filtered high-quality data via SFT. Since SFT can
only imitate the filtered teacher outputs, we further refine the student via RL against $R_{\textsc{Sphere}}$, enabling it to discover spatial configurations beyond the teacher's distribution.

\subsection{Reward Design}
\label{sec:reward_design}
% \ishwarya{This section needs proper motivation of the rewards and citations, and mention of the fact that they are \textit{a singular} choice of possible, valid reward components based on the literature.}
Next, we introduce \textsc{Sphere} (Spatial Heuristic Rewards) for our post-training. The design principles governing our reward suite are as follows: first, it should rank spacious, balanced mixes above narrow, unbalanced ones, following music mixing conventions \cite{pestana14bestpractices}. Moreover, the reward function should reflect human perception of spatial audio aesthetics, with higher-reward data preferred by humans over lower-reward ones.
Additionally, since it is used to calculate advantage scores during RL, it should provide a smooth gradient. In Section~\ref{sec:raw_spatial_metrics} we calculate six raw spatial metrics that quantify mix quality. 
We then convert these raw metrics into verifiable rewards via data-driven thresholding and gaussian and sigmoid smoothing (i.e., reward shaping) in Section~\ref{sec:data_driven_thresholding}.
% \ishwarya{Please add citations to all the component explanations below!}

% Finally, we also want the reward function design to be concise and contain as few hand-tuned hyperparameters as possible.

\subsubsection{Raw Spatial Metrics}
\label{sec:raw_spatial_metrics}
\paragraph{Centering and Spread Metrics} A spatially rich mix should be perceived as centered overall, and with individual stems spread across the panorama \cite{pestana14bestpractices,lopes2023instrument}. To reflect this, the centering and spread metrics are calculated as follows. For each stem $i$, let $\ell_i$ denote its original loudness (in dB LUFS) and recall that $g_i$, $\alpha_i$ are the predicted gain and azimuth respectively. We define the linear-scale weight $w_i = 10^{(\ell_i + g_i)/10}$, reflecting effective
loudness after the gain is applied. For each spatial axis, we measure centering via a gain-weighted mean and spread via an unweighted standard deviation:

  \begin{equation}
  \bar{\alpha}_w = \frac{\sum_i w_i \alpha_i}{\sum_i w_i} \quad
  \sigma_\alpha = \sqrt{\frac{1}{N}\sum_i (\alpha_i - \bar{\alpha})^2}
  \end{equation}

Elevation centering $\bar{\phi}_w$ and spread metrics $\sigma_\phi$ are calculated analogously. The mean is gain-weighted because louder stems dominate perceived centering, while the standard deviation is unweighted because spread should reflect all stem positions equally \cite{blauert1997spatial}. This asymmetry
also prevents potential reward hacking:  $\sigma_\alpha$ cannot be inflated by manipulating gains alone. A desired spatial mix should have both mean metrics near zero (centered image) and both standard deviation metrics large (wide spatial impression).

% \paragraph{Level Balance Metrics}
% Beyond spatial positioning, a good mix maintains balanced loudness across stems. We measure this from 2 aspects: spotlight prominence and per-role level balance.
% Spotlight prominence measures whether the spotlight stem (e.g., vocal, lead guitar) stands out above the rest, while per-role level balance ensures that stems sharing the same musical function (e.g., guitar and piano accompaniment) are at consistent loudness
% levels.

% To quantify these, we first need to know each stem's role in the final mix. Instrument roles, however, are not always available in the metadata $M$. Hence, we rely on the audio language model to assign each stem a role $c_i$ --- one of spotlight, foundation, percussion, accompaniment, or atmosphere --- at inference time. We group stems by their assigned role $c$.

% Spotlight prominence $\Delta_{\mathrm{spot}}$ is the loudness difference (in dB) between the combined spotlight stems and the combined non-spotlight stems.
% For per-role balance, we compute the standard deviation of effective loudness $\sigma_{\ell}^{(c)}$ among stems sharing the same role. Lower $\sigma_{\ell}^{(c)}$ means more consistent levels within a role. Since multiple roles may be active, we aggregate them into $\bar{\sigma}_\ell$ via geometric mean (Section~\ref{sec:continuous_reward_shaping}).

\paragraph{Level Balance Metrics} A good mix should also maintain balanced loudness relationships across stems, and with the spotlight track (e.g., lead vocal) clearly audible \cite{pestana14bestpractices, lopes2023instrument}. To reflect this, we measure two properties: spotlight prominence ($\Delta_{\mathrm{spot}}$) and per-role level consistency ($\sigma_\ell^{(c)}$).
Spotlight prominence is measured as the loudness difference between the spotlight stems and the combination of remaining stems, and per-role level consistency is measured as the loudness's standard deviation among stems sharing the same musical role $c$ (e.g., guitar and piano accompaniment). Both metrics require knowing each stem's role, which is not always available. Hence, we task the ALM with assigning each
stem a role $c$, one of spotlight, foundation, percussion, accompaniment, or atmosphere, as part of its output. A well-balanced mix should have a near-zero spotlight prominence, indicating the spotlight tracks are clearly audible, and low per-role level deviation. The raw metrics, along with their corresponding category and desired direction are summarized in Table \ref{tab:metrics}.

\begin{table}
  \centering
  \resizebox{0.95\columnwidth}{!}{%
  \begin{tabular}{@{}lllll@{}}
  \toprule
  \textbf{Category} & \textbf{Raw Metric} & \textbf{Desired} & \textbf{Smooth.} & \textbf{Symbol} \\
  \midrule
  \multirow{2}{*}{Centering}
    & Azimuth\ mean $\bar{\alpha}_w$             & $\to 0^\circ$  & Gaussian & $r_{\mathrm{az}}^{\mathrm{mean}}$ \\
    & Elevation\ mean $\bar{\phi}_w$               & $\to 0^\circ$  & Gaussian & $r_{\mathrm{el}}^{\mathrm{mean}}$ \\
  \midrule
  \multirow{2}{*}{Spread}
    & Azimuth\ std $\sigma_\alpha$               & large           & Sigmoid  & $r_{\mathrm{az}}^{\mathrm{std}}$ \\
    & Elevation\ std $\sigma_\phi$                 & large           & Sigmoid  & $r_{\mathrm{el}}^{\mathrm{std}}$ \\
  \midrule
  \multirow{2}{*}{Level bal.}
    & Spotlight prom.\ $\Delta_{\mathrm{spot}}$ & $\approx 0$ dB & Gaussian & $r_{\mathrm{spot}}$ \\
    & Per-role std $\sigma_\ell$                & $\to 0$        & Gaussian & $r_{\mathrm{role}}$ \\
  \bottomrule
  \end{tabular}%
  }
  \caption{Summary of sub-rewards in $R_{\textsc{Sphere}}$.}
  \label{tab:metrics}
  \end{table}

\subsubsection{Data-driven Thresholds and Reward Shaping}
\label{sec:data_driven_thresholding}
\label{sec:continuous_reward_shaping}
Given the raw metrics and the desired directions in Table~\ref{tab:metrics}, we further use a function $r$ to convert each raw metric into a bounded sub-reward $\in (0, 1]$, parametrized by cutoff thresholds $\tau$ that distinguish good outputs from poor ones. To avoid hand-selecting thresholds for each metric, which is hard to tune reliably, we propose to derive data-driven thresholds from the teacher model's output distribution. More specifically, for each metric we compute the $p$-th percentile boundary in the desired direction in Table~\ref{tab:metrics}. For instance, for centering metrics, whose desired raw values are near zero, the threshold $\tau$ can be derived as the radius from zero covering the top $p$\% of teacher data points. For spread metrics, higher values indicate wider spatial spread. Hence, $\tau$ is the floor that excludes the narrowest $(1{-}p)$\% of the teacher distribution.

After obtaining $\tau$, the simplest solution is to apply a step function at each threshold ($r = 1$ when the metric satisfies the desired direction in Table~\ref{tab:metrics}, $r = 0$ otherwise).
Although sufficient for SFT filtering, a step function provides sparse gradients for RL, as a sample just below the threshold receives the same zero reward as one far below.
We instead replace the step function with a continuous relaxation, with the data-driven thresholds $\tau$ as the half-reward point: $r(\tau) = 0.5$ for each metric.
Specifically, we apply a Gaussian for closer-to-target metrics and a Sigmoid for higher-is-better metrics in Table~\ref{tab:metrics}:
  \begin{equation}
  r_k = e^{-\frac{k^2}{2\sigma^2}}, \quad k \in \{\bar{\alpha}_w,\, \bar{\phi}_w,\, \Delta_{\mathrm{spot}},\, \sigma_\ell^{(c)}\}
  \end{equation}
  \begin{equation}
  r_k = \frac{1}{1 + e^{-(k - \mu)/\gamma}}, \quad k \in \{\sigma_\alpha,\, \sigma_\phi\}
  \end{equation}

Here, the half-reward constraint $r_k(\tau) = 0.5$ determines the parameters for the reward functions: $\sigma = \tau / \sqrt{2\ln 2}$ for the Gaussian; $\mu = \tau$ and $\gamma = \tau / (3\ln 2)$ for the Sigmoid. Applying these shaping functions yields one sub-reward per metric, besides per-role level balance, since one recording can include stems spanning multiple roles, each with its own $\sigma_\ell^{(c)}$. We aggregate these sub-rewards via
geometric mean over the $K$ active roles, for those that contain more than one stem: $r_{\sigma_\ell} = \left(\prod_{c=1}^{K} r_{\sigma_\ell^{(c)}}\right)^{1/K}$. Here, the geometric mean acts as a soft ``and'': if any single role has high variance, $r_{\sigma_\ell}$ drops sharply, preventing the RL agent from sacrificing one role to optimize others. The total \textsc{Sphere} reward sums all six sub-rewards, for $k \in \{\bar{\alpha}_w,\, \bar{\phi}_w,\, \Delta_{\mathrm{spot}},\, \sigma_\ell^{(c)},\, \sigma_\alpha,\, \sigma_\phi\}$:
\begin{equation}
R_{\textsc{Sphere}} = \sum_{k} r_k
\end{equation}
For readability, Table~\ref{tab:metrics} assigns a named symbol to each sub-reward, which we use in subsequent sections. Notably, sub-reward pairs in \textsc{Sphere} form a non-linear reward landscape, which makes reward optimization non-trivial. For instance, placing all stems at the center maximizes centering rewards but minimizes spread rewards, and vice versa. We later empirically verify that removing any sub-reward pair leads to reward hacking in Section~\ref{sec:results_ablations}, validating the reward design.

\section{Experiments}
\label{sec:experiments}

\subsection{Dataset}
\label{sec:dataset}

We use the MedleyDB multitrack dataset \cite{bittner2014medleydb} for our experiments, which contains 193 full-length songs spanning diverse genres, each provided with a stereo mix reference, isolated mono stems, and a metadata file. We segment each song's mono stems into aligned, non-overlapping 30-second windows, yielding
1650 unique data points. Due to context limitations, we allow a maximum of 29 mono stems (30 seconds each) per data point, resulting in input sequences of up to 20k audio tokens.

% where each data point contains aligned segments from all mono stems and the overall metadata.

Following Section \ref{sec:task} and \ref{sec:paradigm}, we use Gemini 2.5 Pro as the teacher to generate candidate spatial configurations and use Qwen2.5-Omni (3B) as the student model. The prompt includes segmented mono stems and metadata, the stereo reference mix for additional reference, as well as a detailed mixing instruction. The full prompt is provided in Appendix~\ref{app:prompt}.
For each segment, we run five independent inferences to obtain diverse spatial configurations, producing 8243 data points in total (7 failures). 23.5\% of these data points are excluded, which include more than 29 stems, exceeding our student model's context budget. We then perform a song-level train-test split on the remaining 6304 data points
to prevent timbre leakage across segments of the same song, yielding 15 test songs (618 data points) and 139 train songs (5686 data points). The percentile parameter $p$ for data filtering thresholds in Section~\ref{sec:data_driven_thresholding} is set to 0.8. Once the thresholds are defined, we retain only training data points where
all six sub-rewards in Table \ref{tab:metrics} satisfy $r_k \geq \kappa$ and we set $\kappa = 0.5$. This produces 1408 high-quality data points for the student.

\subsection{Training Setup}
\label{sec:training_setup}

During the SFT phase, we finetune the student model on the filtered data via LoRA (rank 8, $\alpha{=}16$) for 20 epochs, distributed across 16 H100 GPUs. The audio encoder is kept trainable. We use a
learning rate of $2{\times}10^{-4}$ with a cosine schedule and 5\% linear warmup. Since each input can contain up to 29 mono stems, we set the max sequence length to 32768, the maximum supported by
the student. Starting from an SFT checkpoint, we further improve the student with GRPO \cite{shao24deepseekmath} for up to 800 steps. Due to the long input sequences, each RL step requires generating and scoring multiple full-length outputs, and an 800-step run takes approximately 24 hours on 8 H100 GPUs.

We employ an asynchronous training
infrastructure that decouples policy optimization from generation: a training engine (6 H100s) updates the model weights while a separate inference engine (2 H100s) generates new spatial
configurations asynchronously, with importance sampling correction to account for the off-policy lag between the two. For the training engine, we apply LoRA (rank 8, $\alpha{=}16$)
to all attention and MLP projections while freezing the audio encoder. We use PPO-style clipping with $\epsilon{=}0.2$, and no KL penalty \cite{Schulman17PPO}.
For the generation engine, at each step, a mini-batch of 6 input segments is sampled and the student generates $G{=}4$ spatial configurations per segment for GRPO, using temperature 1.0 and top-p=0.95.

\subsection{Baselines and Ablations}
\label{sec:baselines}

Due to modality compatibility and sequence length constraints (details in Section \ref{app:limitation}), we only evaluate: 3 proprietary models (Gemini~2.5~Pro, 2.5~Flash and 2.0~Flash), and 5 open-source models (Qwen2.5-Omni-3B, 7B, and Qwen3-Omni Instruct, Captioner, and Thinking).
All baselines are evaluated zero-shot on the same 618 test data points with identical prompts.

% Since our task requires audio language models (ALM) to support multiple audio inputs within a single prompt and have sufficient context length to accommodate up to 29 mono stems (Section~\ref{sec:dataset},~\ref{sec:intro}),
% For example, Gemma 3N models \cite{Gemma3} accept only a single audio input per prompt and Qwen2-Audio has context length only up to 8192 tokens. Among the models that meet both criteria,

To study the effect of data quality versus quantity during SFT, we vary the filtering threshold $\kappa \in \{0.2, 0.3, 0.4, 0.5 \text{ (ours)}\}$ where a data point is accepted if all six sub-rewards satisfy $r_k \geq \kappa$.
This yields 4074, 3152, 2283, and 1408 filtered data points respectively. To validate our reward design during RL, we conduct ablations along two axes. First, we replace the continuous reward shaping (Section~\ref{sec:data_driven_thresholding})
with a binary step function to verify whether continuous rewards provide a more effective learning signal for RL. Second, we disable pairs of sub-rewards corresponding to each category in Table~\ref{tab:metrics} (i.e., spatial centering, spread and level balance)
and analyze the training dynamics.
% (1)~without spatial spread
% ($r_{\mathrm{az}}^{\mathrm{std}}, r_{\mathrm{el}}^{\mathrm{std}}$), (2)~without spatial centering ($r_{\mathrm{az}}^{\mathrm{mean}}, r_{\mathrm{el}}^{\mathrm{mean}}$), and (3)~without level balance
% ($r_{\mathrm{spot}}, r_{\mathrm{role}}$).

\subsection{Evaluation}
\label{sec:evaluation}

\paragraph{Subjective Evaluation}
To validate that \textsc{Sphere} aligns with human preference, we conduct an AB listening test on teacher-generated mixes. We select 10 segments from the test set spanning diverse genres; for each, the
teacher is prompted multiple times, from which we select a high- and low-reward mix with $R_{\textsc{Sphere}}$ gap of at least 1.0. We then render these 20 mixes using the spatializer described
in Section~\ref{sec:task} and loudness-normalized to $-23$~LUFS for fair comparison. In each trial, participants are presented with the two mixes in randomized order and indicate which spatial mix they prefer. Each trial takes 26 minutes on average.

We recruit 89 participants and apply a three-stage screening protocol to ensure reliable responses: headphone verification \cite{Huggingpitch}, gap detection \cite{musiek2005gin}, and a spatial perception
check (distinguish binaural from mono audio, identify which mix is more centered, and has better level balance).
Of 89 participants, 44 passed all three stages (10 trials each, 440 total). The primary causes of exclusion were headphone verification failure (24\%) and gap detection (21\%).

We analyze the results with three statistical tests: (1) a one-sided binomial test on the overall preference rate across pooled trials, (2) a one-sided Wilcoxon signed-rank test on per-listener preference rates and (3) a paired Wilcoxon signed-rank test comparing $R_{\textsc{Sphere}}$ scores of preferred versus non-preferred mixes.

After post-training, we have additionally conducted a listening test that compares 4 systems: student, teacher, and our post-trained models (SFT, SFT+RL). We randomly sample 10 data points from our test set and render binaural upmixes from these 4 systems. We then recruit 10 expert listeners with music or spatial audio backgrounds following the same filtering protocol as above, and conduct a reference-free multi-stimulus preference test. 
Specifically, each participant is presented with the same track, but rendered from these 4 systems in random order. They are then asked to rate their preferences on a continuous 0-10 scale, with 0 and 10 indicating least and most preferred respectively.

\paragraph{Objective Evaluation}

We evaluate all models on the 618 held-out test data points, reporting $R_{\textsc{Sphere}}$, all its sub-rewards and the format pass rate (whether the model produces valid JSON).
To ensure our results reflect stable and representative performance, we report averaged SFT results over the last 100 training steps, where performance has converged. For RL, since the exploration is stochastic,
we run each configuration 3 times independently and report the mean and standard deviation of the highest reward across runs. We additionally plot all sub-rewards against training steps to analyze convergence and potential reward hacking.

\section{Results}
\label{sec:results}

\subsection{Reward-Human Preference Alignment}
\label{sec:results_subjective}

\begin{figure}[t]
\centering
\includegraphics[width=0.8\columnwidth]{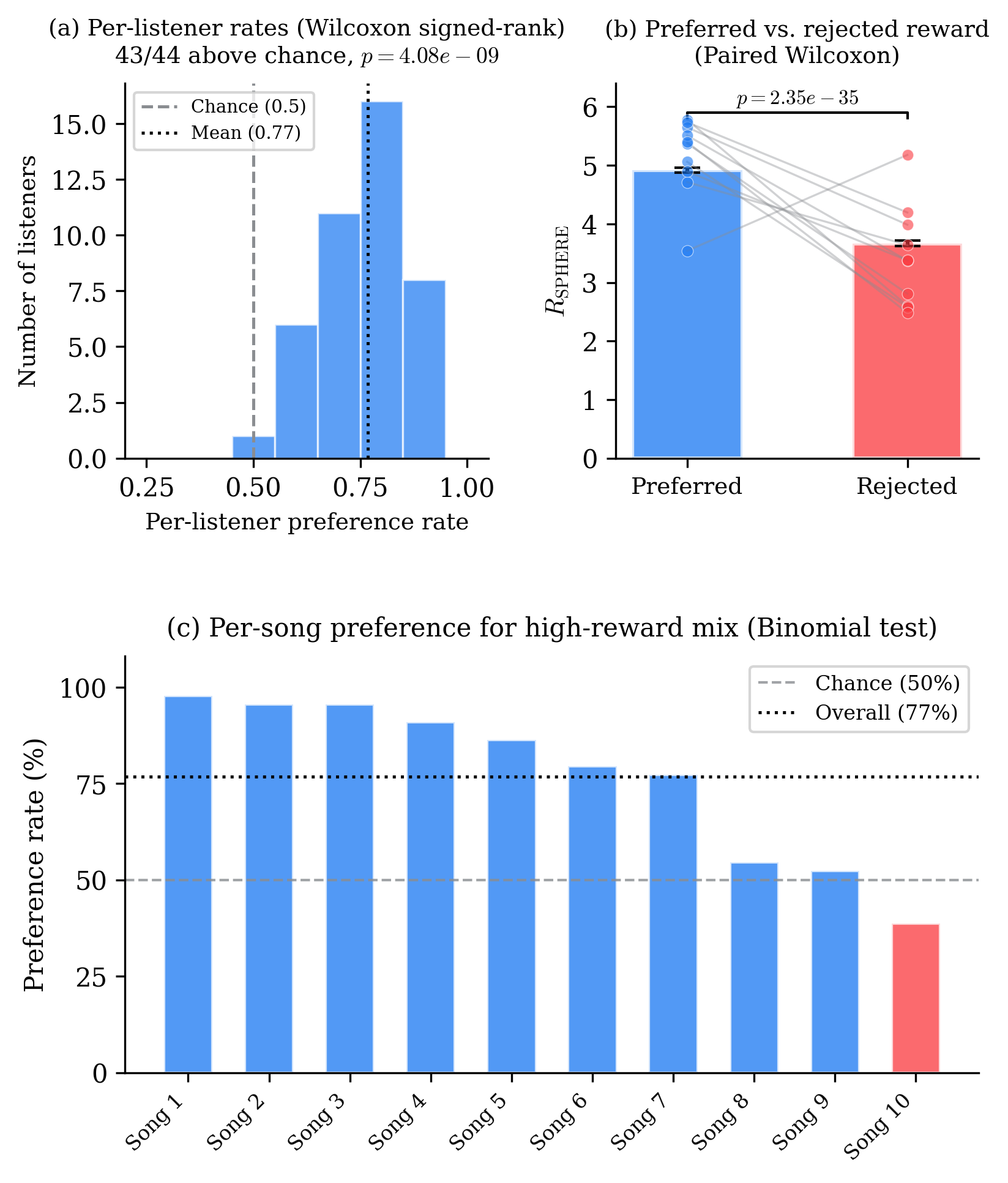}
\caption{AB Listening test results ($N{=}44$ valid participants, 10 trials each). \textbf{(b)}~Mean $R_{\textsc{Sphere}}$ of listener-preferred vs.\ listener-rejected mixes; gray lines connect paired scores for each song (10 pairs).}
\label{fig:preference_reward}
\end{figure}

We first verify whether \textsc{Sphere} aligns with human preference and Figure~\ref{fig:preference_reward} summarizes the results. Overall, the binomial test result (Figure~\ref{fig:preference_reward} c) shows that listeners preferred the high-reward mix in 338 out of 440 trials (76.8\%), which is significantly above chance, with $p = 6.65 \times 10^{-31}$.
Moreover, the one-sided Wilcoxon signed-rank test confirms that individual preference rates are significantly above chance ($p = 4.08 \times 10^{-9}$, $r = 0.79$), with 43 out of 44 listeners
individually preferring the high-reward mix (mean preference rate: $0.77 \pm 0.11$).
Furthermore, in the paired Wilcoxon signed-rank test, mixes preferred by listeners had significantly higher $R_{\textsc{Sphere}}$ than the not-preferred mixes ($4.92 \pm 0.88$ vs.\ $3.67 \pm 1.01$, $p = 2.35 \times 10^{-35}$). These results confirm that \textsc{Sphere} is a meaningful proxy for human spatial audio preference, justifying the use of \textsc{Sphere} as both the data-quality filter and the RL objective.

\subsection{Post-Training Effectiveness}
\label{sec:results_post_training}
\begin{table*}[t]
\centering
\small
\begin{tabular}{@{}lcccccccc@{}}
\toprule
\textbf{Model} & $r_{\mathrm{az}}^{\mathrm{mean}}$ & $r_{\mathrm{az}}^{\mathrm{std}}$ & $r_{\mathrm{el}}^{\mathrm{mean}}$ & $r_{\mathrm{el}}^{\mathrm{std}}$ & $r_{\mathrm{spot}}$ & $r_{\mathrm{role}}$ & \textbf{Fmt\%} & \textbf{Total} \\
\midrule
\multicolumn{9}{@{}l}{\textit{Baselines}} \\
Gemini 2.5 Pro (Teacher) & 0.797 & 0.705 & 0.729 & 0.849 & 0.744 & 0.724 & 100.0 & 4.548 \\
Gemini 2.5 Flash\textsuperscript{\dag} & 0.741 & 0.609 & 0.673 & 0.708 & 0.653 & 0.577 & 98.3 & 3.961 \\
Gemini 2.0 Flash\textsuperscript{\dag} & 0.589 & 0.645 & 0.524 & 0.855 & 0.907 & 0.513 & 95.6 & 4.033 \\
Qwen2.5-Omni-3B (Student) & 0.918 & 0.177 & 0.956 & 0.182 & 0.802 & 0.539 & 80.1 & 3.574 \\
Qwen2.5-Omni-7B & 0.855 & 0.270 & 0.900 & 0.309 & 0.810 & 0.485 & 94.2 & 3.630 \\
Qwen3-Omni-30B-A3B-Instruct & 0.743 & 0.580 & 0.679 & 0.731 & 0.784 & 0.664 & 98.9 & 4.180 \\
Qwen3-Omni-30B-A3B-Captioner & 0.623 & 0.549 & 0.611 & 0.749 & 0.664 & 0.437 & 98.4 & 3.632 \\
Qwen3-Omni-30B-A3B-Thinking\textsuperscript{\ddag} & 0.677 & 0.602 & 0.795 & 0.652 & 0.775 & 0.778 & 41.7 & 4.278 \\
\midrule
\multicolumn{9}{@{}l}{\textit{SFT Phase (avg.\ over last 100 training steps)}} \\
SFT ($\kappa{=}0.2$) & 0.751 & 0.668 & 0.797 & 0.810 & 0.694 & 0.796 & 100.0 & 4.517$\pm$.060 \\
SFT ($\kappa{=}0.3$) & 0.781 & 0.656 & 0.800 & 0.815 & 0.731 & 0.822 & 99.8 & 4.605$\pm$.018 \\
SFT ($\kappa{=}0.4$) & 0.765 & 0.659 & 0.797 & 0.825 & 0.744 & 0.819 & 99.9 & 4.609$\pm$.052 \\
SFT ($\kappa{=}0.5$, ours) & 0.754 & 0.682 & 0.798 & 0.825 & 0.774 & 0.824 & 98.8 & 4.657$\pm$.032 \\
\midrule
\multicolumn{9}{@{}l}{\textit{RL Phase (each avg.\ over 3 separate runs)}} \\
SFT + RL (w/o smoothing) & 0.777 & 0.713 & 0.813 & 0.844 & 0.781 & 0.848 & 99.7 & 4.775$\pm$.070 \\
SFT + RL (w/o $r_{\mathrm{az}}^{\mathrm{std}}, r_{\mathrm{el}}^{\mathrm{std}}$) & 0.947 & 0.275 & 0.955 & 0.401 & 0.863 & 0.948 & 100.0 & 4.388$\pm$.088 \\
SFT + RL (w/o $r_{\mathrm{az}}^{\mathrm{mean}}, r_{\mathrm{el}}^{\mathrm{mean}}$) & 0.681 & 0.787 & 0.733 & 0.896 & 0.823 & 0.862 & 99.9 & 4.782$\pm$.080 \\
SFT + RL (w/o $r_{\mathrm{spot}}, r_{\mathrm{role}}$) & 0.806 & 0.740 & 0.799 & 0.859 & 0.778 & 0.809 & 99.8 & 4.791$\pm$.057 \\
SFT + RL (ours) & 0.816 & 0.759 & 0.734 & 0.890 & 0.879 & 0.903 & 100.0 & \textbf{4.981$\pm$.013} \\
\bottomrule
\end{tabular}
\caption{Objective evaluation results. Each $r$ represents a sub-reward in $R_{\textsc{Sphere}}$ (higher is better). Fmt\% is the percentage of outputs with valid formatting. $\kappa$ is minimum per-metric threshold for SFT filtering. \textsuperscript{\dag}Evaluated on N=573; 45 items with 29 or more stems were excluded due to API failure. \textsuperscript{\ddag}Low format pass rate: chain-of-thought reasoning exhausts the context window.}
\label{tab:results}
\end{table*}
% \ishwarya{Include figure showing qualitative results - parameter predictions (plotted on sphere) for a small list of stem names as inputs, etc.}
Table~\ref{tab:results} summarizes the objective evaluation results. Among zero-shot baselines, Gemini 2.5 Pro achieves the highest $R_{\textsc{Sphere}} = 4.548$, followed by Qwen3-Omni-Instruct
at 4.180; Qwen2 models trail behind. We observe two notable failure patterns. First, Qwen2.5-Omni-3B (our student) scores the highest centering ($r_{\mathrm{az}}^{\mathrm{mean}} = 0.918$) but the
lowest spread ($r_{\mathrm{az}}^{\mathrm{std}} = 0.177$), indicating that the model defaults to placing all sources at the center, achieving high centering at the expense of spatial diversity. Second,
Qwen3-Omni-Thinking achieves a competitive $R_{\textsc{Sphere}} = 4.278$ on valid outputs, but only 41.7\% of its responses pass formatting: since our input contains multiple audio stems that can
occupy over 20k of the 32k context window, the reasoning consumes the remaining budget before the model produces a spatial configuration. These two failure modes (i.e., collapsed spatial
behavior and context-length exhaustion) illustrate the difficulty of zero-shot binauralization.

After SFT on reward-filtered teacher data, the same 3B student (SFT with $\kappa=0.5$) reaches
4.657, even surpassing the teacher. RL further improves the total reward to $4.981$, the highest score across all models. The full pipeline yields a gain of +1.407 over the untrained student, and even
surpasses the much larger teacher model by a large margin, demonstrating that post-training with \textsc{Sphere} is effective even on a small open-source model.

% \ishwarya{We are giving 3-digit precision metric values here; how significant are these improvements? Can we do analysis to map reward changes to perceptual significance?  This may be a rabbit hole now, but reviewers will likely ask.}
We next study per-metric changes to understand how SFT and RL individually improve the model. During SFT, the student first learns to produce valid outputs, with format pass rate
improving from 80.1\% to 98.8\%. SFT also corrects the student's degenerate behavior (placing all sources at the center): centering metrics $r_{\mathrm{az}}^{\mathrm{mean}}$ and $r_{\mathrm{el}}^{\mathrm{mean}}$ decrease from 0.918 to 0.754 and 0.956 to 0.798 respectively.
Meanwhile, spread metrics $r_{\mathrm{az}}^{\mathrm{std}}$ and $r_{\mathrm{el}}^{\mathrm{std}}$ increase dramatically, from 0.177 to 0.682 and 0.182 to 0.825 respectively. Moreover, level balance $r_{\mathrm{role}}$ also improves from 0.539 to 0.824.

RL further refines the student by targeting the weaker metrics after SFT by exploring the output space. The largest gains are in spotlight prominence $r_{\mathrm{spot}}$, which improves from 0.774 to 0.879, and level balance $r_{\mathrm{role}}$ improves from 0.824 to 0.903. SFT left these two sub-rewards the most room for improvements. After RL, spread metrics further improve, from 0.682 to 0.759 for $r_{\mathrm{az}}^{\mathrm{std}}$, and 0.825 to 0.890 for $r_{\mathrm{el}}^{\mathrm{std}}$.
These are at the cost of a slight decrease in elevation centering: $r_{\mathrm{el}}^{\mathrm{mean}}$ drops from $0.798$ to $0.734$. RL also resolves the remaining formatting failures, resulting in 100\% correct format rate. In summary, SFT improves the student
by imitating teacher behavior, but remains bounded by the quality of the teacher data. RL further improves this by reinforcing the model's own high-reward outputs and shifts the output distribution towards higher-quality spatial configurations.

\subsection{Ablation Studies}
\label{sec:results_ablations}

\begin{figure*}[t]
\centering
\includegraphics[width=0.9\textwidth]{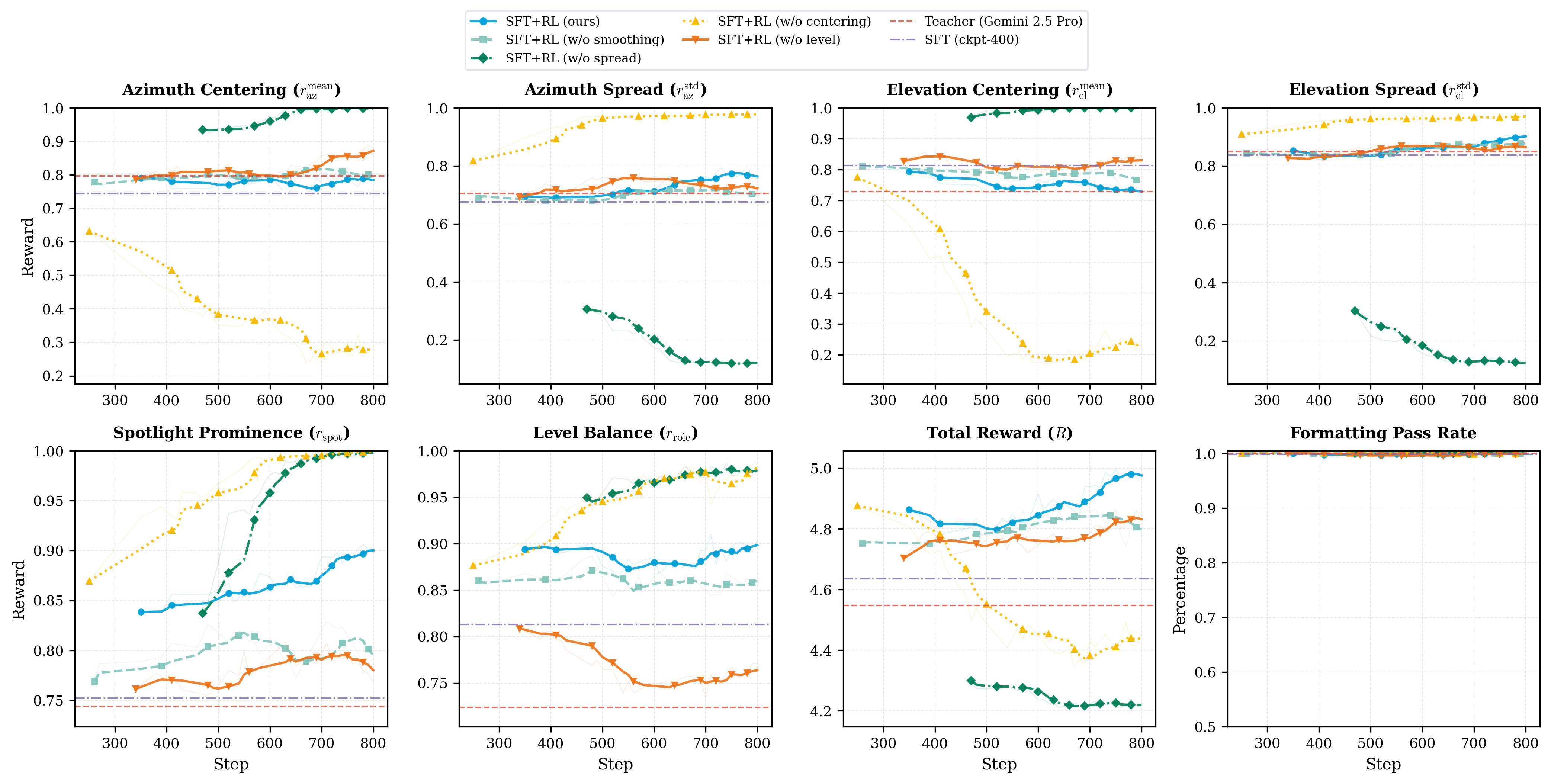}
\caption{RL training dynamics. Removing individual sub-reward pairs leads to reward hacking on remaining metrics. RL with continuous relaxation outperforms the step-function baseline.}
\label{fig:rl_training_curves}
\end{figure*}

Results of our ablation studies are shown in Table~\ref{tab:results} and Figure \ref{fig:rl_training_curves}. We report 2 findings: (1) data quality is more important than quantity during SFT and (2) removing sub-reward pairs can lead to reward hacking during RL. For the SFT data filtering ablation (Table~\ref{tab:results}), stricter filtering ($\kappa = 0.5$) retains only 1408 data points, yet achieves the highest $R_{\textsc{Sphere}}$ of 4.657, outperforming $\kappa = 0.2$, which uses nearly 3 times more data, yet reaches only 4.517. More broadly, $R_{\textsc{Sphere}}$ increases monotonically with $\kappa$: from 4.517 at $\kappa = 0.2$, to 4.605 at $\kappa = 0.3$, 4.609 at $\kappa = 0.4$, and 4.657 at $\kappa = 0.5$. This confirms that data quality matters more than quantity for SFT.

Next, we ablate individual pairs of RL sub-rewards according to Table \ref{tab:metrics}. Figure \ref{fig:rl_training_curves} plots the sub-rewards over training steps for each configuration.
Notably, removing any pair of sub-rewards leads to reward hacking, with severity depending on the removed components. The most severe case is removing the spatial spread sub-rewards ($r_{\mathrm{az}}^{\mathrm{std}}, r_{\mathrm{el}}^{\mathrm{std}}$). Without the incentive to spread sources apart,
the model (labeled in green) learns to place all stems at the center to maximize centering, but spread metrics collapse. This leads to the lowest total reward of 4.388 among all RL configurations. Removing the centering sub-rewards ($r_{\mathrm{az}}^{\mathrm{mean}}, r_{\mathrm{el}}^{\mathrm{mean}}$) produces the mirror effect: the model (labeled in yellow) pushes stems apart to maximize spread, with centering
dropping to 0.681 while spread inflates to 0.787. In both cases, the model exploits the remaining sub-reward at the expense of its removed counterparts. Removing the level balance sub-rewards ($r_{\mathrm{spot}}, r_{\mathrm{role}}$) does not result in severe reward hacking, as shown in the model labeled in orange, but spotlight prominence and per-role level balance both degrade, while the remaining sub-rewards stay stable.
Reward hacking is possible in this case too, where the model could place a few loudest stems in the middle, while scattering quieter stems to inflate spread sub-rewards.
However, this would require the model to deviate significantly from the SFT distribution, and we argue that PPO clipping limits how far the policy can shift per step, potentially preventing dramatic reward hacking.
Finally, when all 6 sub-rewards are present (model labeled in darker blue), $R_{\textsc{Sphere}}$ achieves the highest total with no metric collapsing, suggesting that the reward combination is resistant to hacking. Moreover, replacing continuous reward shaping with a binary step function (model labeled in lighter blue) reduces total reward from 4.981 to 4.775, with noticeably higher variance across runs in Table \ref{tab:results}.
This confirms that the Gaussian and sigmoid shaping introduced in Section~\ref{sec:data_driven_thresholding} provides a more effective learning signal for RL.

\subsection{Subjective Evaluation of Post-Trained Models}

  \begin{table}[th]
  \centering
  \small
  \begin{tabular}{@{}lcc@{}}
  \toprule
  \textbf{Model} & \textbf{Mean} & \textbf{Std} \\
  \midrule
  Qwen2.5-Omni-3B (Student) & 3.24 & 2.20 \\
  Gemini 2.5 Pro (Teacher)  & 5.55 & 2.80 \\
  \midrule
  SFT (ours)                       & 5.71 & 2.74 \\
  SFT+RL (ours)                 & 5.73 & 2.85 \\
  \bottomrule
  \end{tabular}
  \caption{Listening test results that compare models before and after post-training (10 expert listeners,
  10 tracks). Preference ratings on a continuous
  0 to 10 scale, where 10 is most preferred.}
  \label{tab:mushra}
  \end{table}

We further compare our post-trained models with the student and teacher model via a multi-stimulus listening test. The results in Table \ref{tab:mushra} show that the student model is the least preferred. After SFT and RL post-training, the overall ratings are improved to 5.71 and 5.73, outperforming the teacher model, with a rating of 5.55. Interestingly, although the listening test result follows the same trend as the objective $R_{\textsc{Sphere}}$ in Table~\ref{tab:results} (SFT+RL ranks highest, followed by SFT, the teacher, and the student), we are seeing smaller improvements from SFT to SFT+RL (5.71 to 5.73), compared to the SFT phase (3.24 to 5.71). This shows potential further directions when designing reward suites for post-training: each sub-reward might have a different perceptual weight (e.g., spotlight prominence might affect the overall preference more easily than centering metrics). We leave a systematic study of perceptually-weighted rewards to future work.

\section{Limitations and Future Directions}
  \label{app:limitation}

While effective for automatic music binaural upmixing, our approach has two main limitations.
First, it is constrained by modality compatibility, as most existing frontier models have limited support for the audio modality. Among major frontier model families as of 2026 (e.g., Claude, GPT, Gemini, DeepSeek, and Qwen), only Gemini, Qwen, Gemma, and GPT-4o support audio input. Among these, GPT-4o and Gemma3N are primarily designed for speech understanding, while Gemma3N supports only a single audio input per prompt. Second, context length is a main bottleneck. The Qwen models tokenize audio at 25 tokens per second \cite{xu2025qwen25omni}. A single 30-second stem thus consumes around 750 tokens. With prompt overheads, the 32k-token context window accommodates at most 29 30-second stems, yet real-world multitrack recordings often exceed this limit and contain full-length audio beyond 30 seconds. This caused the exclusion of 23.5\% of our dataset in Section \ref{sec:dataset}. This large context window also makes both SFT and RL training expensive. Future work could explore context-length scaling techniques or efficient tokenizers to accommodate multitrack recordings.

\section{Conclusion}

We presented an approach to automatic music binaural upmixing via ALM post-training, guided by \textsc{Sphere}, a reward inspired by music mixing conventions. We empirically verify that \textsc{Sphere} aligns with human preference and through post-training, our student model surpasses frontier ALMs. Our ablation studies show that SFT data quality is more important than quantity, and all 6 sub-rewards are necessary to prevent reward hacking during RL. More broadly, our work suggests that musical domain expertise can be distilled into reward functions to guide post-training, enabling optimization of ALM outputs with respect to common practices.

% ALMs to improve through exploration while respecting common practices.
%\bibliography{custom,anthology-overleaf-1,anthology-overleaf-2} , and that continuous reward shaping outperforms a step-function baseline for RL

\bibliography{custom}

\appendix
\section*{Appendix}

\section{Prompt Template}
  \label{app:prompt}
The full prompt is shown in Figure~\ref{fig:prompt}. It outputs a per-stem JSON configuration specifying azimuth, elevation, gain, musical role, and a reasoning trace. The prompt structures gain prediction as a three-stage process: gain staging normalization, aesthetic balance, and compensation. Role assignment is performed by the model rather than provided as input, since metadata does not always indicate musical function. Spatial placement principles drawn from mixing conventions \cite{pestana14bestpractices} are included directly in the prompt.

   \begin{figure*}[h]
  \begin{tcblisting}{title=Full Prompt, colback=gray!5, colframe=gray!50, listing only, listing options={basicstyle=\tiny\ttfamily, breaklines=true}}
system_prompt: You are an expert audio engineer and music producer.

user_prompt:
  1. Context & Inputs
  You are a Spatial Audio Mix Engineer. You are provided with:
    1. Metadata: Stem groupings (e.g., S01, S02), individual raw stems (e.g., R01, R02) and component labels (Bass, Melody).
    2. Audio Stems: Raw recordings with corresponding loudness in LUFS.
    3. Reference Stereo Mix: For tonal and balance reference.

  Note on Bleed: Tracks may contain signal leakage (bleed). Your spatial decisions must be based on the intended source of the microphone, not the leaked signals.
  Note on Room Mics: Mute tracks intended to capture room ambience (Gain = -99). Look for keywords like 'Room', 'Ambience', or 'Main System'. Exception: Do not mute 'Overhead' (OH) tracks if they are the primary source for cymbals/high-frequency percussion.
  Note on Empty tracks: Listen to the track segments and identify if empty. Apply gain_db = -99 if the track segment is empty.

  2. The Task

  Step 1 (Overall Big Picture): Identify the genre. Use music mixing knowledge of this genre to guide mixing.
  Step 2 (Analyze & Categorize): Evaluate the role of each stem. Identify each track as: "spotlight" / "foundation" / "percussion" / "accompanying_parts" / "atmosphere" / "others".
  Step 3 (Three-Stage Gain Staging): Calculate gain_db = Gain_A + Offset_B + Offset_C.
    Step A (Normalization): Gain_A = (-18) - (Raw_LUFS)
    Step B (Aesthetic Balance): Apply relative boost/cut based on role.
    Step C (Compensation Gain): Final mental audit of the summed mix.
  Step 4 (Spatial Placement): Assign each stem azimuth and elevation.
  Step 5 (Source Placement Principles):
    - Center Anchor: Lead Vocals, Bass, Kick near center (0 deg azimuth).
    - Clarity through Space: Separate competing frequencies by placing instruments apart from each other.
    - Verticality: Higher-frequency elements at higher elevations.
    - Immersive Depth: Rear quadrants for atmospheric elements.
  Step 6 (Confidence Score): Include confidence score from 0 to 1.

  5. Output Format: Reasoning and JSON.
  {"confidence_score": 0.8,
   "stems": {"S01": {"instrument": "vocal",
     "raw": {"R01": {"filename": "lead_vocal.wav", "track_role": "spotlight",
       "reasoning": "...", "azimuth": 0, "elevation": 0, "gain_db": -6.5}}}}}
  \end{tcblisting}
  \caption{Full prompt used for teacher generation and student inference.}
  \label{fig:prompt}
  \end{figure*}

\section{SFT Training Dynamics}
  \label{app:sft}

\begin{figure*}
\centering
\includegraphics[width=\textwidth]{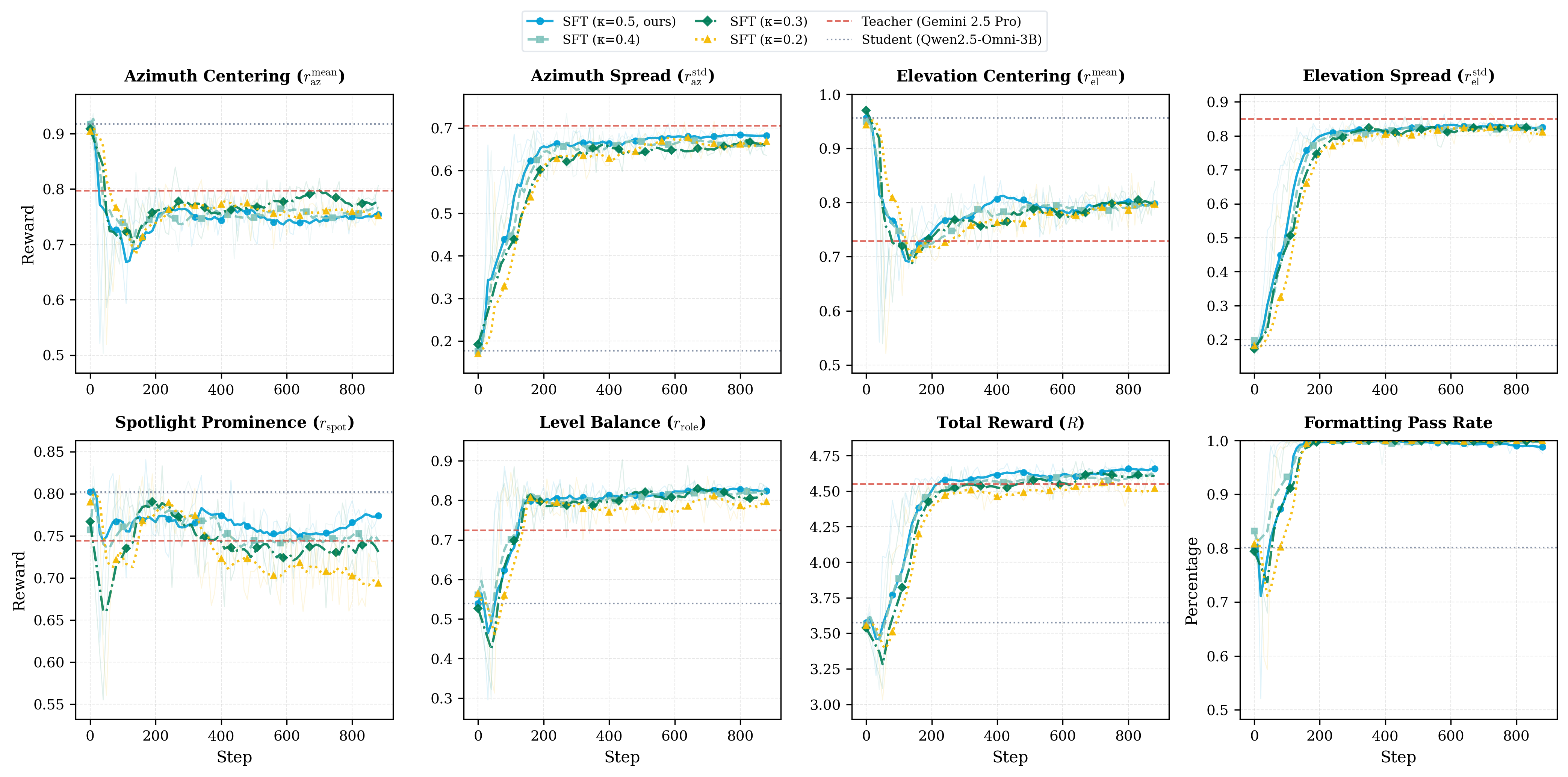}
\caption{SFT training dynamics across filtering thresholds $\kappa \in \{0.2, 0.3, 0.4, 0.5\}$, with stricter filtering ($\kappa = 0.5$) achieving the highest total reward.}
\label{fig:sft_curves}
\end{figure*}

Figure~\ref{fig:sft_curves} plots the reward dynamics on the test set throughout SFT for each filtering threshold $\kappa$. Three phases are visible across all configurations. In the first $\sim$100 steps, the model rapidly learns output formatting and corrects its degenerate zero-shot behavior: azimuth and elevation spread climb sharply from their collapsed baselines, while centering metrics decrease as the model begins distributing stems across the panorama rather than clustering them at center. Between steps $\sim$100-200 all sub-rewards quickly increase and slowly converge towards their final reward value after step $\sim$200.

The $\kappa = 0.5$ model achieves the highest total reward despite using the fewest training samples, confirming the data-quality-over-quantity finding reported in Section~\ref{sec:results_ablations}. The red dashed line (teacher performance) is surpassed by all SFT configurations on total reward, demonstrating that even basic reward filtering enables the student to exceed the teacher.

% Between steps 100--400, all sub-rewards converge toward their final values. The main differences emerge in spotlight prominence ($r_{\mathrm{spot}}$) and level balance ($r_{\mathrm{role}}$), where stricter filtering ($\kappa = 0.5$) produces consistently higher and more stable rewards.

% After step 400, all configurations plateau.
% Overall speaking, all models start from the student distribution and transition to the teacher distribution, labeled in grey and red respectively, with the model with the highest data quality ($\kappa$=0.5) achieving the highest reward.

\section{Label-to-Role Assignment Analysis}
  \label{app:role}

As discussed in Section \ref{sec:raw_spatial_metrics}, the level balance sub-rewards rely on roles predicted by the ALM, raising the question of whether 
these assignments are reliable. Since no ground truth exists for the full role taxonomy, direct evaluation is difficult. However, the MedleyDB dataset provides 
two component labels (i.e., melody and bass), which we use for partial validation and analysis. Table~\ref{tab:role} reports label-to-role assignment distribution, mapping MedleyDB's ground-truth labels to predicted track roles. Bass $\to$ Fnd.\ measures the percentage of bass-labeled stems assigned \texttt{foundation}. For stems labeled as melody, we report the percentage assigned \texttt{spotlight}, \texttt{accompanying\_parts}, or other roles.

Table~\ref{tab:role} shows that bass stems are assigned \texttt{foundation} most of the time by all ALMs, whereas melody-labeled stems are primarily
assigned as \texttt{spotlight} or \texttt{accompanying\_parts} by ALMs. This reveals a fundamental challenge in music semantics research: existing dataset annotations are static and song-level, whereas musical roles are inherently time-variant and context-dependent. For instance, a jazz piece: Oil, in the MedleyDB dataset has both the piano and guitar stems labeled as \texttt{melody}. Yet in a given segment the guitar solos while the piano comps. The ALM correctly assigns them \texttt{spotlight} and \texttt{accompanying\_parts} respectively. In another classical piece (MusicDelta Beethoven), all instruments (flute, clarinet, oboe, trumpet, French horn) share the metadata label: melody, yet ALMs assign them to \texttt{others}, which is also semantically correct since they do not belong to any suitable roles we defined in Section \ref{sec:raw_spatial_metrics}.

Despite these challenges, Table~\ref{tab:role} indicates our post-trained model assigns roles based on segment-level musical context rather than defaulting to metadata-driven shortcuts (e.g., always assigning \texttt{spotlight} to stems labeled \texttt{melody} in the prompt).  Our analysis also reveals a fundamental challenge: text descriptions of music are usually static, whereas 
we empirically show that music is inherently context-dependent, and song-level description would lead to information bottlenecks. We believe this gap
   presents an opportunity for future work in multimodal music representation learning.

\begin{table}[h]
\centering
\resizebox{0.95\columnwidth}{!}{%
\begin{tabular}{@{}lcccc@{}}
\toprule
\textbf{Model} & \textbf{Bass$\to$Fnd.} & \textbf{Mel$\to$Spot.} & \textbf{Mel$\to$Acc.} & \textbf{Mel$\to$Oth.} \\
\midrule
\multicolumn{5}{@{}l}{\textit{Baselines}} \\
Gemini 2.5 Pro (Teacher) & 97.9 & 30.7 & 53.1 & 16.2 \\
Gemini 2.5 Flash & 100.0 & 31.4 & 67.6 & 0.9 \\
Gemini 2.0 Flash & 94.2 & 12.1 & 54.0 & 33.9 \\
Qwen2.5-Omni-3B (Student) & 93.9 & 30.1 & 50.4 & 19.4 \\
Qwen2.5-Omni-7B & 93.9 & 29.6 & 24.1 & 46.3 \\
Qwen3-Omni-Instruct & 97.1 & 25.3 & 72.2 & 2.5 \\
Qwen3-Omni-Captioner & 99.7 & 33.2 & 41.3 & 25.5 \\
Qwen3-Omni-Thinking & 100.0 & 39.6 & 57.5 & 2.8 \\
\midrule
\multicolumn{5}{@{}l}{\textit{Post-trained}} \\
SFT ($\kappa{=}0.5$, ours) & 99.3 & 46.2 & 51.3 & 2.5 \\
SFT + RL (ours) & 99.7 & 26.9 & 59.5 & 13.6 \\
\bottomrule
\end{tabular}%
}
\caption{Label-to-Role assignment analysis using MedleyDB \texttt{component} labels and ALM predicted roles. MedleyDB's \texttt{component} label is a static song-level label whereas roles assigned by ALM are segment-level.}
\label{tab:role}
\end{table}

\end{document}